\documentclass[12pt,a4paper]{article}
\usepackage[top=25truemm,bottom=35truemm,left=25truemm,right=25truemm]{geometry}
\usepackage{amsmath,amssymb}
\usepackage{graphicx}

\begin{document}

\setlength{\baselineskip}{18pt}
\begin{titlepage}
\begin{flushright}
\begin{tabular}{l}
 SU-HET-10-2014\\
 EPHOU-14-014
\end{tabular} 
\end{flushright}

\vspace*{1.2cm}
\begin{center}
{\Large\bf Gravitational effects on vanishing Higgs potential at the Planck scale
}
\end{center}
\lineskip .75em
\vskip 1.5cm

\begin{center}
{\large Naoyuki Haba}$^1$,
{\large Kunio Kaneta}$^2$,
{\large Ryo Takahashi}$^1$, and
{\large Yuya Yamaguchi}$^3$\\

\vspace{1cm}

$^1${\it Graduate School of Science and Engineering, Shimane University,\\
 Matsue 690-8504, Japan}\\
 $^2${\it ICRR, University of Tokyo,\\
 Kashiwa, Chiba 277-8582, Japan}\\
$^3${\it Department of Physics, Faculty of Science, Hokkaido University,\\
 Sapporo 060-0810, Japan}\\

\vspace{10mm}
{\bf Abstract}\\[5mm]
{\parbox{13cm}{\hspace{5mm}
 
We investigate gravitational effects on the so-called multiple point criticality principle (MPCP)
 at the Planck scale.
The MPCP requires two degenerate vacua,
 whose necessary conditions are expressed by vanishing Higgs quartic coupling $\lambda(M_{\rm Pl})=0$
 and vanishing its $\beta$ function $\beta_\lambda(M_{\rm Pl})=0$.
We discuss a case that a specific form of gravitational corrections are assumed
 to contribute to $\beta$ functions of coupling constants
 although it is accepted that gravitational corrections do not alter
 the running of the standard model (SM) couplings.
To satisfy the above two boundary conditions at the Planck scale,
 we find that the top pole mass and the Higgs mass should be
 170.8\,GeV$\lesssim M_t\lesssim$171.7\,GeV and $M_h=125.7\pm0.4$\,GeV, respectively,
 as well as include suitable magnitude of gravitational effects (a coefficient of gravitational contribution as $|a_\lambda| > 2$).
In this case, however, since the Higgs quartic coupling $\lambda$ becomes negative below the Planck scale,
 two vacua are not degenerate.
We find that $M_h \gtrsim 131.5$\,GeV with $M_t \gtrsim 174$\,GeV
 is required by the realization of the MPCP.
Therefore, the MPCP at the Planck scale
 cannot be realized in the SM and also the SM with gravity
 since $M_h \gtrsim 131.5$\,GeV is experimentally ruled out.

}}
\end{center}
\end{titlepage}

\section{INTRODUCTION}
The ATLAS \cite{Aad:2012tfa} and CMS \cite{Chatrchyan:2012ufa} Collaborations
 of the Large Hadron Collider (LHC) experiment observed the Standard Model (SM) like the Higgs boson,
 whose pole mass $M_h$ is obtained by \cite{Giardino:2013bma}
\begin{eqnarray}
	M_h = 125.7 \pm 0.4\,{\rm GeV}.
\label{Higgs_mass}
\end{eqnarray}
The LHC data are almost consistent with the SM predictions,
 and there are no signs of new physics beyond the SM at present.
In addition, the observed values of Higgs and top masses suggest the SM can be valid up to the Planck scale
 in light of the vacuum (meta) stability.
The multiple point criticality principle (MPCP)  requires two degenerate vacua between EW and Planck scales~\cite{Froggatt:1995rt}.
%
Necessary conditions of the MPCP (NC-MPCP) at the Planck scale is expressed
 by vanishing the Higgs quartic coupling $\lambda(M_{\rm Pl})=0$
 and its $\beta$ function $\beta_\lambda(M_{\rm Pl})=0$.
It was pointed out that these boundary conditions (BCs) suggest
 $135\pm9$\,GeV Higgs mass with the top pole mass as $173\pm5$\,GeV~\cite{Froggatt:1995rt}.
Reference \cite{Shaposhnikov:2009pv} showed that the Higgs mass of 126.5\,GeV
 with the top mass of 171.2\,GeV implies both $\lambda(M_{\rm Pl})\simeq0$
 and $\beta_\lambda(M_{\rm Pl})\simeq0$ in a scenario of asymptotic safety of gravity
 (see also \cite{Holthausen:2011aa}-\cite{Branchina:2013jra} for more recent analyses).
Note that the MPCP could be realized at ${\cal O} (10^{17})$\,GeV
 with the use of a lighter magnitude of the top mass as 171\,GeV without gravitational effects.\footnote{
If one considers the realization of the MPCP at a lower energy scale than the Planck scale,
 the desired values of the top mass are small compared to the MPCP at the Planck scale
 as $M_t=171$ (146) GeV for $\lambda(\mu)=\beta_\lambda(\mu)=0$
 with $\mu=10^{18}$ ($10^{10}$) GeV~\cite{Froggatt:1995rt}
 (e.g., see also Ref.~\cite{Buttazzo:2013uya} for more recent analyses).}

Since a gravitational interaction becomes important near the Planck scale,
 we must take into account gravitational effects for discussions around the Planck scale.
However, general relativity is a nonrenormalizable theory by perturbation methods,
 and the theory of quantum gravity has not been established yet.
Nevertheless, treating general relativity as an effective field theory provides a practical method 
 to find the effects of quantum gravity \cite{Donoghue:1993eb, Donoghue:1994dn}.
Because of the effective theory approach,
 gravitational contributions to the renormalization group equation (RGE) evolution of couplings
 are calculated by using different regularization schemes
 \cite{Robinson:2005fj}-\cite{Zanusso:2009bs}.
Since the MPCP at the Planck scale predicts close values of the Higgs and top masses
 to the experimental ones,
 it might be important for the SM with gravity to accurately analyze gravitational effects
 with the latest experimental results.
We discuss a case that a specific form of gravitational corrections are assumed
 to contribute to $\beta$ functions of coupling constants
 (although it is accepted that gravitational corrections do not alter the running of the standard model couplings).

In this paper,
we will first clarify regions of model parameters satisfying the NC-MPCP.
To satisfy the NC-MPCP, we will find that the top pole mass should be
 in the region of 170.8\,GeV$\lesssim M_t\lesssim$171.7\,GeV
 with the Higgs mass as $M_h=125.7\pm0.4$\,GeV, and typical magnitude of the gravitational effects. 
Moreover, numerical analyses will show that the NC-MPCP can be typically satisfied
 in region of a coefficient of gravitational contribution to $\beta_\lambda$ as $|a_\lambda|>2$
 with negative gravitational contributions to $\beta$ functions of the top Yukawa and gauge couplings.
In this case, however, since the Higgs quartic coupling $\lambda$ becomes negative below the Planck scale,
 two vacua are not degenerate.
Then, we will find that $M_h \gtrsim 131.5$\,GeV with $M_t \gtrsim 174$\,GeV
 is required by the realization of the MPCP.
Therefore, the MPCP cannot be realized in the SM and also the SM with gravity
 since $M_h \gtrsim 131.5$\,GeV is experimentally ruled out.

\section{STANDARD MODEL WITH GRAVITATIONAL EFFECTS}

We discuss gravitational effects on the SM in this section. In particular, we 
show the effects on $\beta$ functions of the SM coupling constants and the 
resultant evolution of the couplings under the RGEs.

\subsection{Gravitational effects on {\boldmath$\beta$} functions of coupling constants}

We consider the SM and quantized general relativity,
 in which the graviton couples with the SM particles.
The gravitational coupling constant is given by $\kappa^2 = 16\pi G = 16\pi/M_{\rm Pl}^2$,
 in which $G$ is the Newton constant, and the Planck mass is
 $M_{\rm Pl}=G^{-1/2}\simeq1.22\times10^{19}$\,GeV.
This coupling is small enough to neglect far below the Planck scale,
 while it is effective around the Planck scale.
To see energy dependences of couplings,
 we have to solve the corresponding RGEs.
The RGE for a coupling constant $x$ is given by
\begin{eqnarray}
	\frac{dx}{d\ln \mu} = \beta_x = \beta_x^{\rm SM} + \beta_x^{\rm gravity},
 \label{rge}
\end{eqnarray}
 where $\beta_x^{\rm SM}$ and $\beta_x^{\rm gravity}$ are the Callan-Symanzik 
$\beta$ function in the SM and a gravitational contribution to the $\beta$ functions, respectively.\footnote{
$\beta$ functions in the SM at the two-loop level are given in the Appendix.
In our numerical analysis, we utilize the SM $\beta$ functions at the two-loop level
 and include the leading contributions from gravity given in Eq.\,(\ref{b}).}
The gravitational contribution at the one-loop level is written by 
\begin{eqnarray}
	\beta_x^{\rm gravity} &=& \frac{a_x x}{(4\pi)^2} \kappa^2 \mu^2
		= \frac{a_x}{\pi} \frac{\mu^2}{M_{\rm Pl}^2}\, x,
 \label{b}
\end{eqnarray}
 where $a_x$ is a constant that denotes the gravitational contribution to the 
 coupling $x$, and it can be obtained by calculating the corresponding graviton one-loop diagrams.

In this paper, we assume $a_{g_1} = a_{g_2} = a_{g_3} \equiv a_{g}$ for the SM gauge
 couplings as taken in~\cite{Shaposhnikov:2009pv}.
This assumption seems to be valid due to the universality of the gravitational interactions.
If $a_g$ is negative, all gauge couplings approach zero near the Planck scale.
Thus, the gauge interactions are asymptotically free.
In addition, the Landau pole problem for the $U(1)_Y$ gauge coupling in the SM can be solved
 due to the presence of the fixed point. 
References~\cite{Robinson:2005fj}-\cite{He:2010mt} obtained negative values of 
$a_g$ with $|a_g|\sim{\cal O}(1)$.
For the top Yukawa coupling $y_t$,
 the Yukawa interaction also becomes asymptotically free when $a_{y_t}<0$.
Regarding the Higgs quartic coupling $\lambda$, 
 the gravitational contribution $\beta_\lambda^{\rm gravity}$ can be dominant around the Planck scale.
Thus, the vacuum stability ($\lambda > 0$ up to the Planck scale)
 can be more easily realized compared to the SM 
by taking the gravitational contribution with $a_\lambda>0$ into account.
In fact, $a_\lambda = 6$ was utilized in Refs.\,\cite{He:2010mt, Wang:2013lzq}.
We note that Refs.~\cite{Pietrykowski:2006xy,Toms:2011zza} and \cite{Zanusso:2009bs} showed that
 the values of $a_{g}$ and $a_{y_t}$ depend on the gauge fixing, respectively.
In addition, Ref.~\cite{Anber:2011ut} showed that
 gravitational effects of the gravitational constant depend on physical processes.
Therefore, we treat $a_g$, $a_{y_t}$ and $a_\lambda$ as free parameters in this work.

\subsection{Gravitational effects on RGE evolution}
Next, we show typical gravitational effects on the RGE evolution of coupling constants. 
To solve the RGEs, the following BCs \cite{Buttazzo:2013uya, Hamada:2014xka} are taken,
\begin{eqnarray}
	&&g_Y(M_t) = 0.35761 + 0.00011 \left( \frac{M_t}{{\rm GeV}} - 173.10 \right), \qquad g_1 = \sqrt{\frac{5}{3}}g_Y, \label{g1_mt}\\
	&&g_2(M_t) = 0.64822 + 0.00004 \left( \frac{M_t}{{\rm GeV}} - 173.10 \right),\\
	&&g_3(M_t) = 1.1666 - 0.00046 \left( \frac{M_t}{{\rm GeV}} - 173.10 \right) + 0.00314 \left( \frac{\alpha_s(M_Z) - 0.1184}{0.0007} \right),\\
	&&y_t(M_t) = 0.93558 + 0.00550 \left( \frac{M_t}{{\rm GeV}} - 173.10 \right) -0.00042 \left( \frac{\alpha_s(M_Z) - 0.1184}{0.0007} \right), \label{yt_mt}\\
	&&\lambda(M_t) = 0.12711 - 0.00004 \left( \frac{M_t}{{\rm GeV}} - 173.10 \right) + 0.00206 \left( \frac{M_h}{{\rm GeV}} - 125.66 \right), \label{lambda_mt} \\
	&&\alpha_s(M_Z) = 0.1184 \pm 0.0007,
\end{eqnarray}
 where $M_t$ and $M_h$ are the pole masses of the top quark and the Higgs boson, respectively.
The BCs of the gauge and top Yukawa couplings are 
determined by Eqs.\,(\ref{g1_mt})-(\ref{yt_mt}) for given $M_t$ and $\alpha_s$. 
After solving the RGEs, Eq.\,(\ref{lambda_mt}) gives $M_h$ as
\begin{eqnarray}
M_h = 125.66 + \frac{1}{0.00206} \left[ \lambda(M_t) - 0.12711 + 0.00004 \left( \frac{M_t}{{\rm GeV}} - 173.10 \right) \right]\, {\rm GeV}.
\end{eqnarray}

\begin{figure}[t]
  \begin{center}
      \begin{minipage}{0.45\hsize}
        \begin{center}
(a)\\
          \includegraphics[clip, width=\hsize]{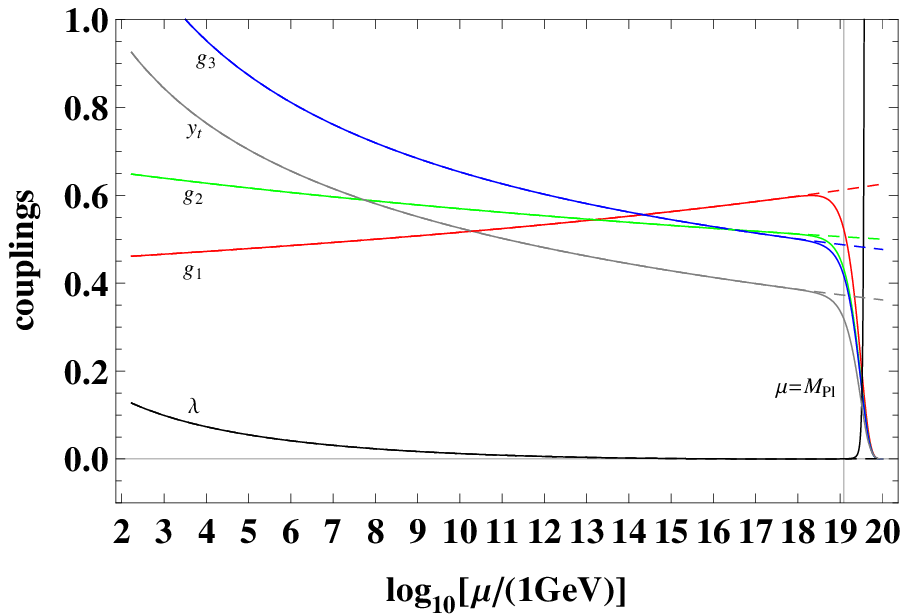}
        \end{center}
      \end{minipage}\hspace{5mm}
      \begin{minipage}{0.45\hsize}
        \begin{center}
(b)\\
          \includegraphics[clip, width=\hsize]{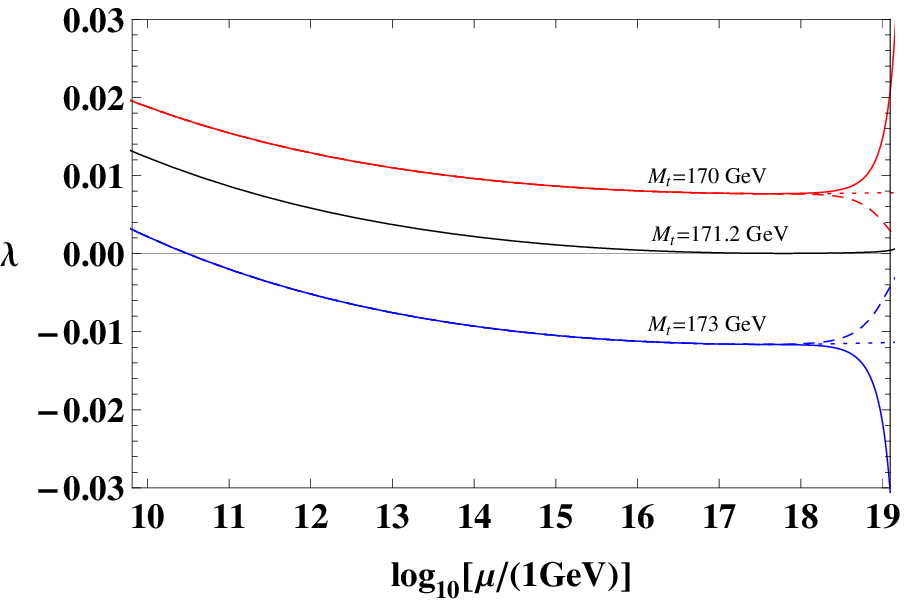}
        \end{center}
      \end{minipage}
  \end{center}
   \caption{(a) Typical RGE evolution of the SM couplings.
The solid (dashed) lines show the RGE evolution with (without) the gravitational contributions.
We take $a_g=a_{y_t}=-1$, $a_\lambda=6$, $M_t=171.2$\,GeV, 
$M_h=125.7$\,GeV, and $\alpha_s=0.1184$.
(b) $a_\lambda$ and $M_t$ dependences of the evolution of $\lambda$.
We take $a_g=a_{y_t}=-1$, $M_h=125.7$\,GeV, and $\alpha_s=0.1184$.
The solid, dashed, and dotted lines correspond to $a_\lambda = 6$, $a_\lambda = -6$, and $a_\lambda = 0$, respectively.
The red, black, and blue lines are the cases of $M_t=170$, 171.2, and 173\,GeV, respectively.
}
	\label{couplings}
\end{figure}

In Fig.\,\ref{couplings}, we show an example of the RGE evolution of the SM couplings
 that include gravitational effects,
 in which the above BCs are taken.
Figure~\ref{couplings} (a) shows typical RGE evolution of the SM couplings.
The solid (dashed) lines show the RGE evolution with (without) the gravitational contributions. 
In the figure, we take $a_g=a_{y_t}=-1$ \cite{Shaposhnikov:2009pv,Tang:2008ah},
 $a_\lambda=6$ \cite{He:2010mt,Wang:2013lzq},
 $M_t=171.2$\,GeV, $M_h=125.7$\,GeV, and $\alpha_s=0.1184$,
 which realize $\lambda(M_{\rm Pl})=0$.
The RGE evolution of gauge and top Yukawa couplings indicates 
 asymptotic freedom of the interactions due to the gravitational contributions as mentioned above.
If we take $a_g>0$ and $a_{y_t}>0$,
 the gauge and top Yukawa couplings become large around the Planck scale.
One can see that the gravitational contributions become effective above $\mu \approx 10^{18}$\,GeV.
In this work, we concentrate on the energy region up to $\mu=M_{\rm Pl}$ as a physical scale,
 which is depicted by the vertical solid line.
On the other hand, Ref.\,\cite{He:2010mt} mentioned that
 the analyses can be valid in the region of $|x^{-1}\beta_x^{\rm gravity}|<1$,
 which reads $\mu < (\pi/|a_x|)M_{\rm Pl}$.
In this case, an effective energy region for the model is larger than the Planck scale when $|a_x|<\pi$.
As a result, the gravitational effects with $a_g=-1$ may assist the gauge coupling unification
 since all gauge couplings unify at an energy scale lower than $\mu<\pi M_{\rm Pl}$.

Figure\,\ref{couplings} (b) shows the $a_\lambda$ and $M_t$ dependence of the evolution of $\lambda$.
We take $a_g=a_{y_t}=-1$, $M_h=125.7$\,GeV, and $\alpha_s=0.1184$ in the figure.
The solid, dashed, and dotted lines correspond to $a_\lambda = 6$, $a_\lambda = -6$, and $a_\lambda = 0$, respectively.
The red, black, and blue lines are the cases of $M_t=170$\,GeV, 171.2\,GeV, and 173\,GeV, respectively.
Note that the sign of $\lambda$ in the Planck scale strongly depends on the top pole mass,
 e.g. $\lambda(M_{\rm Pl}) \gtrless 0$ for $M_t \lessgtr 171.2$\,GeV
 when we take $M_h=125.7$\,GeV and $\alpha_s=0.1184$.
One can understand the behaviors of $\lambda$ around the Planck scale by considering $\beta_\lambda$ as follows.
Now $\beta_\lambda$ can be approximately written as
\begin{eqnarray}
	&&\beta_{\lambda}  \approx \frac{1}{(4\pi)^2} \left[ \lambda \left( 12 y_t^2 - 9 g_2^2 \right)
		- 6 y_t^4 + \frac{9}{8}g_2^4 + \frac{9}{20}g_1^2 g_2^2 \right]
		+ \frac{a_\lambda \lambda}{(4\pi)^2} \kappa^2 \mu^2,
\label{b2}
\end{eqnarray}
 around the Planck scale.
When both $a_g$ and $a_{y_t}$ are negative,
 the gauge and top Yukawa couplings approach zero as $\mu$ becomes large.
Then, the last term (the gravitational contribution term) of Eq.~(\ref{b2}) dominates $\beta_\lambda$
 when $a_\lambda$ is sufficiently large.
Thus, $\lambda$ is approximately given by
\begin{eqnarray}
	\lambda \sim \lambda_* \exp \left[ \frac{a_\lambda}{2(4 \pi)^2} \kappa^2 (\mu^2 - \mu_*^2) \right],
\label{lambda_sim}
\end{eqnarray}
 where $\mu_*$ is an energy scale at which the gravitational contribution becomes dominant,
 and $\lambda_*$ is a value of $\lambda$ at $\mu = \mu_*$.
As a result, $|\lambda|$ is larger (lower) for $a_\lambda >0$ ($a_\lambda<0$)
 while keeping the sign of $\lambda$.

One can also see that the gravitational effects and the value of $a_\lambda$ slightly change
 the RGE evolutions of the SM couplings around the Planck scale,
while they significantly change the $\beta$ functions.
It should be remarked that 
$\beta_\lambda(M_{\rm Pl}) = 0$ can be induced from $\lambda(M_{\rm Pl})\simeq0$
 when the gravitational contributions are taken into account in the SM as seen in Eq.~(\ref{b2}). 
Thus, two independent BCs of $\lambda(M_{\rm Pl}) = 0$ and $\beta_\lambda(M_{\rm Pl}) = 0$
 predict almost the same value of the Higgs mass for a given $M_t$.
In the next section, we quantitatively clarify regions of model parameters
 for the realization of the MPCP in this scenario.

\section{NUMERICAL ANALYSES}

In this section, first we will clarify regions of the model parameters
 satisfying the NC-MPCP $\lambda(M_{\rm Pl}) = 0$ and $\beta_\lambda(M_{\rm Pl}) = 0$.
Next, we will discuss whether those parameter regions realize the MPCP,
 which requires two degenerate vacua.

\subsection{Realization of the NC-MPCP}

\begin{figure}[t]
  \begin{center}
      \begin{minipage}{0.48\hsize}
        \begin{center}
          \includegraphics[clip, width=\hsize]{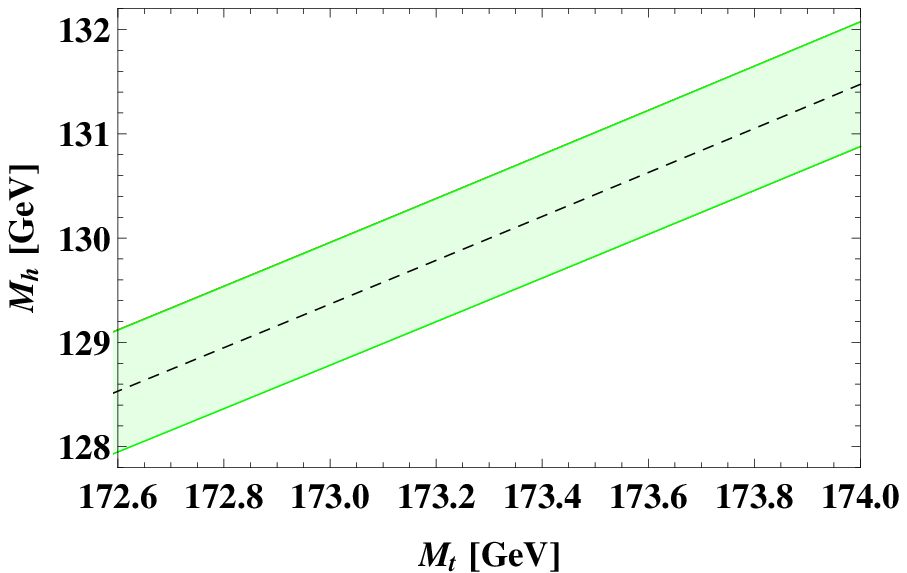}
        \end{center}
      \end{minipage}
\hspace{3mm}
      \begin{minipage}{0.48\hsize}
        \begin{center}
          \includegraphics[clip, width=\hsize]{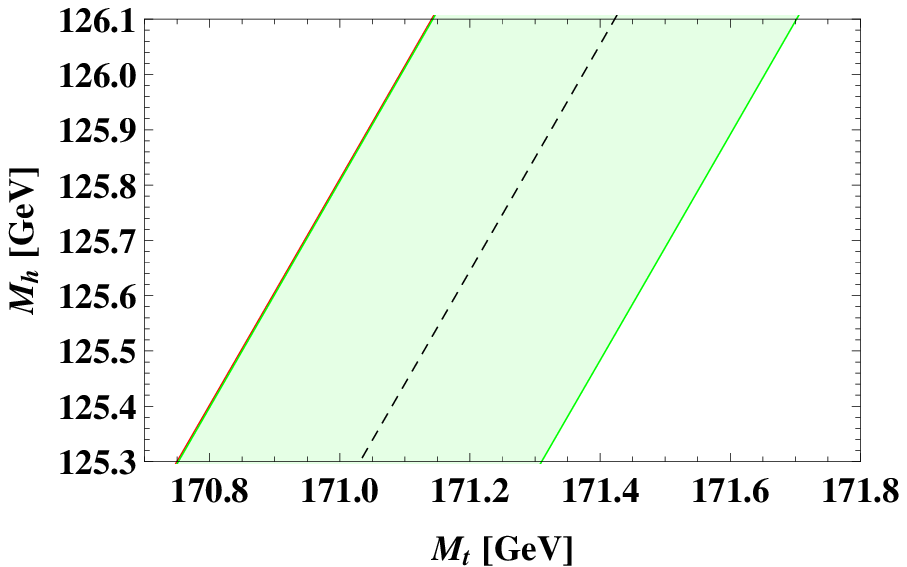}
        \end{center}
      \end{minipage}
  \end{center}
   \caption{Predictions for the Higgs and the top pole masses from the realizations of
 $\lambda(M_{\rm Pl}) = 0$ and $\beta_\lambda(M_{\rm Pl}) = 0$.
Two figures show different numerical regions in the ($M_t$\,[GeV], $M_h$\,[GeV]) plane. 
We take $a_g=a_{y_t}=-1$ and $a_\lambda=6$.
The upper and lower bounds in both figures are given by $\alpha_s = 0.1177$ and $0.1191$, respectively.
The black-dashed line is given by $\alpha_s = 0.1184$.}
	\label{mhmt}
\end{figure}

\begin{figure}[!ht]
  \begin{center}
\qquad (a)\qquad \qquad \qquad \qquad \qquad \qquad(b)\qquad \qquad \qquad \qquad \qquad \qquad (c)\\
          \includegraphics[clip, scale=0.41]{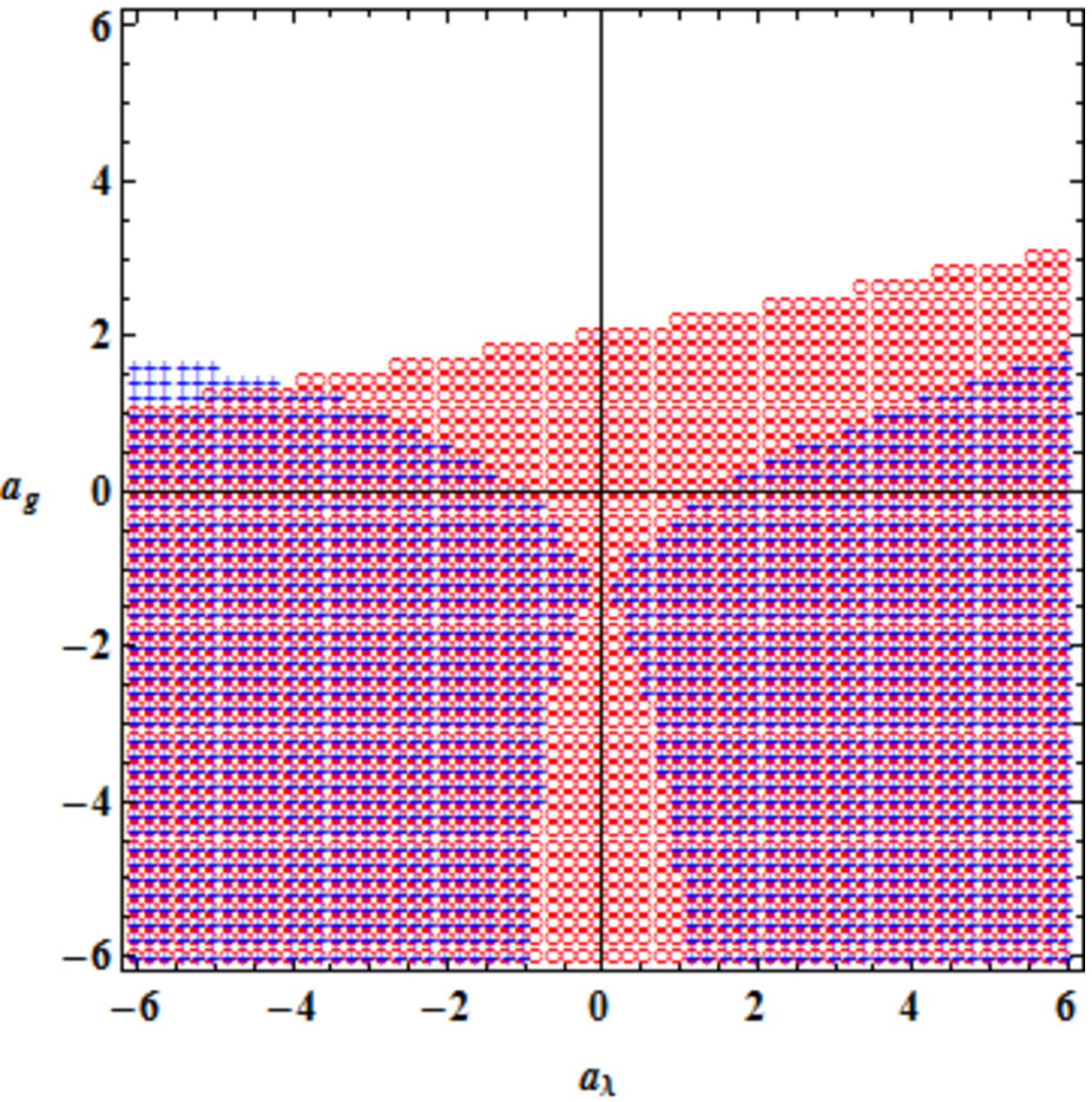}
          \includegraphics[clip, scale=0.41]{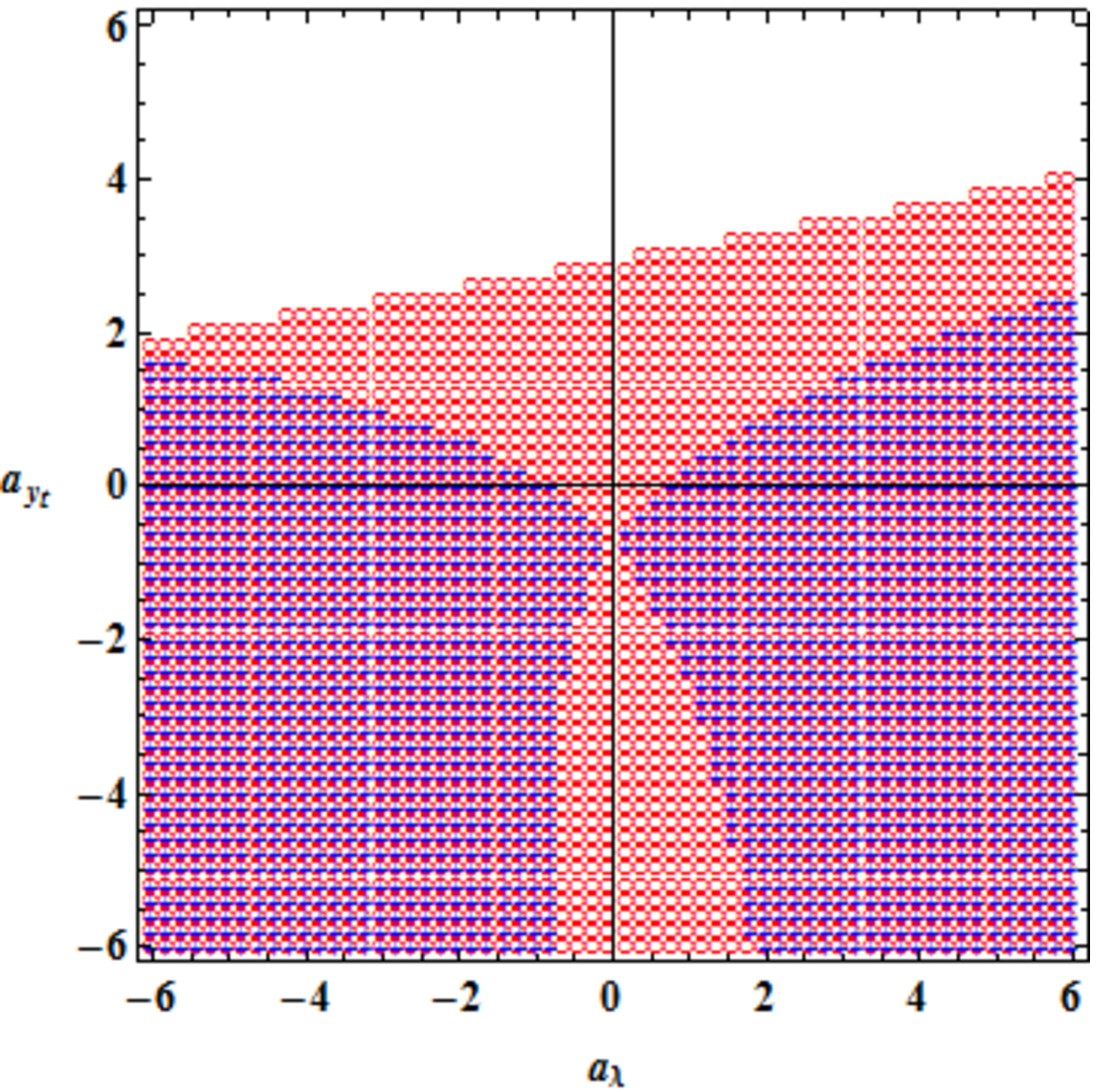}
          \includegraphics[clip, scale=0.41]{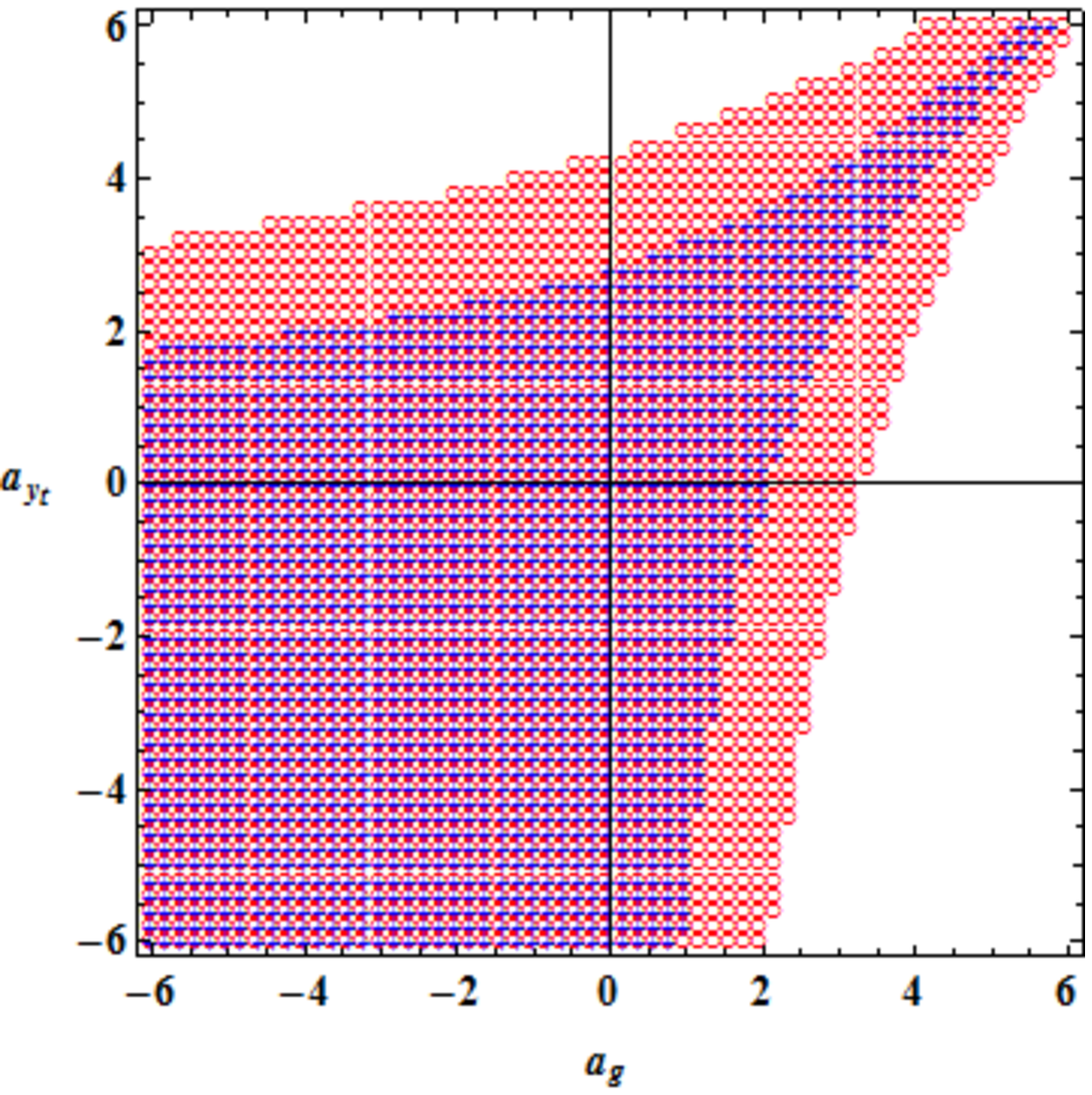}
  \end{center}
  \caption{
Parameter dependence of the realizations of $\lambda(M_{\rm Pl}) = 0$ and $\beta_\lambda(M_{\rm Pl}) = 0$,
 which are shown in (a) ($a_\lambda, a_g$), (b) ($a_\lambda, a_{y_t})$, and (c) ($a_g, a_{y_t}$) planes.
The red circles and blue crosses in all figures indicate 
 $\lambda(M_{\rm Pl}) = 0$ and $\beta_\lambda(M_{\rm Pl}) = 0$, respectively.
We take $\alpha_s = 0.1184$ and $M_t = 171.2$\,GeV in all figure, and 
$a_{y_t}=-1$ in (a), $a_g=-1$ in (b), and $a_\lambda=6$ in (c).
The Higgs pole mass can be within $M_h=125.7\pm0.4~{\rm GeV}$ at all points.}
\label{dependence}
\end{figure}

We show the results of numerical analyses satisfying the NC-MPCP in Figs.~\ref{mhmt} and \ref{dependence}.
Figure \ref{mhmt} shows predictions for the Higgs and the top pole masses from the realizations of 
$\lambda(M_{\rm Pl}) = 0$ and $\beta_\lambda(M_{\rm Pl}) = 0$.
In the figure, we take $a_g=a_{y_t}=-1$ and $a_\lambda=6$.
To realize $M_h = 125.7 \pm 0.4$\,GeV,
 the top pole mass should be in the region of $170.8~{\rm GeV}\lesssim M_t \lesssim 171.7~{\rm GeV}$.\footnote{
On the other hand, the experimentally favored value of the top quark mass is $173.34 \pm 0.76$\,GeV \cite{ATLAS:2014wva}.
In Ref.\,\cite{Hamada:2014xka}, the authors suggest this discrepancy might not be problematic,
 and just the discrepancy between the pole mass and the mass obtained at the collider experiments \cite{ATLAS:2014wva}.
Our predicted value is a pole of a colored quark,
 while the experimentally obtained value is an invariant mass of the color singlet final states.
Since the observed $t \bar{t}$ pair is dominantly color octet at the hadron collider,
 the discrepancy of a few GeV from the singlet final states may be caused \cite{Horiguchi:2013wra,Smith:1996xz}.}
The result is consistent with the previous result including gravitational effects~\cite{Shaposhnikov:2009pv}.


We also show the parameter dependence of the realizations of
 $\lambda(M_{\rm Pl}) = 0$ and $\beta_\lambda(M_{\rm Pl}) = 0$
 for a typical top mass as 171.2\,GeV in Fig.\,\ref{dependence}.
As mentioned above, $M_t = 171.2$\,GeV cannot realize the MPCP
 although the NC-MPCP is satisfied.
The red circles and blue crosses in Fig.~\ref{dependence} indicate the necessary conditions of the MPCP,
 $\lambda(M_{\rm Pl})=0$ and $\beta_\lambda(M_{\rm Pl})=0$, respectively.
Hence, the NC-MPCP can be satisfied at the overlapped points of red circles and blue crosses in the figure.

Let us mention the results of parameter dependence for the NC-MPCP.
First, the NC-MPCP can be satisfied in almost all regions of $a_g<0$ and $a_{y_t}<0$
 [see Figs.~\ref{dependence} (a) and \ref{dependence} (b)]. 
Since gauge and top Yukawa couplings can be asymptotically free in this regime,
 $\beta_\lambda$ is dominated by the gravitational contribution.
As a result, $\beta_\lambda(M_{\rm Pl})=0$ is induced from $\lambda(M_{\rm Pl})\simeq0$ as mentioned above.

Second, regarding the value of $a_\lambda$ for the NC-MPCP, 
 sufficiently large value of $|a_\lambda|$ such as $|a_\lambda| \gtrsim 2$ is typically required 
 [see Figs.~\ref{dependence} (a) and \ref{dependence} (b)].
When $|a_\lambda|$ is sufficiently large,  the gravitational term is dominant in $\beta_\lambda$.
Thus, the conditions $\lambda(M_{\rm Pl})=0$ and $\beta_\lambda(M_{\rm Pl})=0$ are satisfied at the same time.
Note that $\beta_\lambda(M_{\rm Pl}) = 0$ cannot be realized for $-0.001 \lesssim a_\lambda \lesssim 0.04$.

Third, when $a_g$ is positive, gauge couplings become larger as $\mu$ becomes large.
Since the terms positively contribute to $\beta_\lambda$,
 the condition $\beta_\lambda(M_{\rm Pl}) = 0$ leads $\lambda(M_{\rm Pl}) < 0$.
Similarly, top Yukawa coupling becomes larger for $a_{y_t} > 0$,
 and the terms negatively contribute to $\beta_\lambda$.
Then, the condition $\beta_\lambda(M_{\rm Pl}) = 0$ leads $\lambda(M_{\rm Pl}) > 0$.
As a result, for $a_g>0$ ($a_{y_t}>0$) the NC-MPCP cannot be satisfied in most cases.
However, the NC-MPCP can be satisfied when both $a_g$ and $a_{y_t}$ are positive,
 because the contributions of gauge and top Yukawa couplings cancel each other
 [see upper right region in the first quadrant of Fig. 3 (c)].

Finally, there are also regions satisfying the NC-MPCP,
 in which $a_g\gtrless0$ and $a_{y_t}\lessgtr$
 [e.g., see the second and fourth quadrants of Fig. 3 (c)],
 when both $|a_g|$ and $|a_{y_t}|$ are relatively small with sufficiently large $|a_\lambda|$.
Since the last term of Eq. (11) can still dominate the other terms also in the regions,
 $\beta_\lambda(M_{\rm Pl})=0$ can be induced from $\lambda(M_{\rm Pl})\simeq0$.

\subsection{Realization of the MPCP}

\begin{figure}[t]
  \begin{center}
      \begin{minipage}{0.7\hsize}
        \begin{center}
          \includegraphics[clip, width=\hsize]{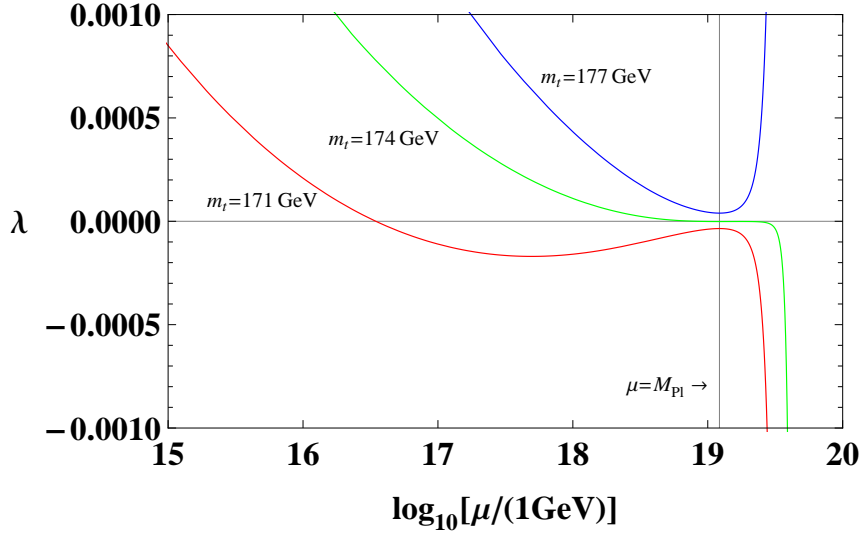}
        \end{center}
      \end{minipage}
  \end{center}
  \caption{RGE evolution of $\lambda$ satisfying $\beta_\lambda(M_{\rm Pl}) = 0$.
We take $a_g=a_{y_t}=-1$, $a_\lambda=6$ and $\alpha_s = 0.1184$.
The red, green and blue lines correspond to $M_t = 171$\,GeV, 174\,GeV and 177\,GeV, respectively.}
	\label{beta0}
\end{figure}

In the previous subsection,
 we investigate the regions of the model parameters satisfying the NC-MPCP.
Figure \ref{beta0} shows examples of the RGE evolution of $\lambda$ satisfying $\beta_\lambda(M_{\rm Pl}) = 0$.
We take $a_g=a_{y_t}=-1$, $M_h=125.7$\,GeV, and $\alpha_s=0.1184$ in the figure.
The red, green and blue lines correspond to $M_t = 171$\,GeV, 174\,GeV and 177\,GeV, respectively.
The behavior around the Planck scale is understood by Eq.~(\ref{lambda_sim}),
 that is, $|\lambda|$ is larger for $a_\lambda >0$ as keeping the sign of $\lambda$.
This figure implies that the Higgs potential has a local maximum at the Planck scale for $M_t \lesssim 174$\,GeV,
 and thus the MPCP cannot be realized for $M_t \lesssim 174$\,GeV.
We find that $M_h \gtrsim 131.5$\,GeV with $M_t \gtrsim 174$\,GeV is required by the realization of the MPCP.
Since the gravitational contribution is effective for $\mu \gtrsim 10^{18}$\,GeV,
 these values of the Higgs and the top masses for the MPCP in our scenario
 are not drastically changed from the ones predicted by the past research
 without taking into account gravity~\cite{Froggatt:1995rt}
 (see also~\cite{Giardino:2013bma,Holthausen:2011aa}).
Therefore, we conclude that the MPCP cannot be realized in the SM and also the SM with gravity
 since $M_h \gtrsim 131.5$\,GeV is experimentally ruled out.

\section{SUMMARY AND DISCUSSION}
We have investigated the realization of the MPCP at the Planck scale,
 which requires the degenerate SM vacua at the Planck scale,
 in the SM with gravitational effects.
The NC-MPCP is expressed by
 the vanishing Higgs quartic coupling $\lambda(M_{\rm Pl}) = 0$
 and its $\beta$ function $\beta_\lambda(M_{\rm Pl}) = 0$.
The vanishing $\beta$ function can be induced from the vanishing Higgs quartic coupling
 with a suitable magnitude of $|a_\lambda|$ and the asymptotic free of Yukawa and gauge interactions
 due to the gravitational effects around the Planck scale.
To satisfy the NC-MPCP,
 the top and the Higgs pole masses should be in the regions of
 170.8\,GeV$\lesssim M_t\lesssim$171.7\,GeV and $M_h=125.7\pm0.4$ GeV, respectively,
 with $a_g=a_{y_t}=-1$, $a_\lambda=6$.
Moreover, numerical analyses have shown that the NC-MPCP can be typically satisfied
 in the region of $|a_\lambda|>2$ with $a_g<0$ and $a_{y_t}<0$.
In this case, however, since the Higgs quartic coupling $\lambda$ becomes negative below the Planck scale
 two vacua are not degenerate.
We have found that $M_h \gtrsim 131.5$\,GeV with $M_t \gtrsim 174$\,GeV
 is required by the realization of the MPCP.
Therefore, the MPCP cannot be realized in the SM and also the SM with gravity
 since $M_h \gtrsim 131.5$\,GeV is experimentally ruled out.


Finally, we comment on a nonminimal coupling of the Higgs to the Ricci scalar $\xi H^2\mathcal{R}$
 which was not considered in this paper.
This nonminimal interaction is useful in the Higgs inflation \cite{Hamada:2014xka}, \cite{Bezrukov:2007ep}-\cite{Branchina:2014usa},
 and the RGE evolution of the SM couplings might be modified around the Planck scale.
Then, it is expected that the Higgs and top masses and allowed regions of $a_x$ might be also modified.
The analyses will be given in a separate publication.

\subsection*{\centering Acknowledgment} \label{Acknowledgement}
This work is partially supported by the Scientific Grant by Ministry of Education 
and Science, No. 24540272. The works of 
R.T. and Y.Y. are supported by Research Fellowships of the Japan Society for the
 Promotion of Science for Young Scientists
 [Grants No. 
24$\cdot$801 (R.T.) and No. 26$\cdot$2428 (Y.Y.)].

\section*{APPENDIX}
\appendix

\section*{{\boldmath$\beta$} FUNCTIONS IN THE SM} \label{app:RGE}
The RGE of coupling $x$ is given by $dx/d\ln \mu = \beta_x$, in which $\mu$ is a renormalization scale. The $\beta$ functions for coupling constants of the SM 
are given by
\begin{eqnarray}
	&&\beta_{g_1} = \frac{g_1^3}{(4\pi)^2} \left[ \frac{41}{10} \right]
		+ \frac{g_1^3}{(4\pi)^4} \left[ - \frac{17}{10}y_t^2 + \frac{199}{50}g_1^2 + \frac{27}{10}g_2^2 + \frac{44}{5}g_3^2 \right],\\
	&&\beta_{g_2} = \frac{g_2^3}{(4\pi)^2} \left[ -\frac{19}{6} \right]
		+ \frac{g_2^3}{(4\pi)^4} \left[ - \frac{3}{2}y_t^2 + \frac{9}{10}g_1^2 + \frac{35}{6}g_2^2 + 12 g_3^2 \right],\\
	&&\beta_{g_3} = \frac{g_3^3}{(4\pi)^2} \left[ -7 \right]
		+ \frac{g_3^3}{(4\pi)^4} \left[ - 2 y_t^2 + \frac{11}{10}g_1^2 + \frac{9}{2}g_2^2 - 26 g_3^2 \right],\\
	&&\beta_{y_t} = \frac{y_t}{(4\pi)^2} \left[ \frac{9}{2}y_t^2 - \frac{17}{20}g_1^2 - \frac{9}{4}g_2^2 - 8 g_3^2 \right] \nonumber\\
		&&\quad\quad + \frac{y_t}{(4\pi)^4} \left[ y_t^2 \left( -12 y_t^2 + \frac{393}{80}g_1^2 + \frac{225}{16}g_2^2 + 36 g_3^2 - 12 \lambda \right) \right. \nonumber\\
		&&\quad\quad \left. + \frac{1187}{600}g_1^4 - \frac{23}{4}g_2^4 - 108 g_3^4 - \frac{9}{20}g_1^2 g_2^2 + \frac{19}{15}g_1^2 g_3^2 + 9 g_2^2 g_3^2 + 6 \lambda^2 \right],\\
	&&\beta_{\lambda}  = \frac{1}{(4\pi)^2} \left[ \lambda \left( 24 \lambda + 12 y_t^2 - \frac{9}{5}g_1^2 - 9 g_2^2 \right)
		- 6 y_t^4 + \frac{27}{200}g_1^4 + \frac{9}{8}g_2^4 + \frac{9}{20}g_1^2 g_2^2 \right] \nonumber\\
		&&\quad\quad + \frac{1}{(4\pi)^4} \left[ \lambda^2 \left( - 312 \lambda - 144 y_t^2 +\frac{108}{5}g_1^2 + 108 g_2^2\right)
		+ \lambda y_t^2 \left( - 3 y_t^2 + \frac{17}{2}g_1^2 + \frac{45}{2}g_2^2 + 80 g_3^2 \right) \right. \nonumber\\
		&&\quad\quad + \lambda \left( \frac{1887}{200}g_1^4 - \frac{73}{8}g_2^4 + \frac{117}{20}g_1^2 g_2^2 \right)
		+ y_t^4 \left( 30 y_t^2 - \frac{8}{5}g_1^2 - 32 g_3^2 \right) \nonumber\\
		&&\quad\quad \left. + y_t^2 \left( - \frac{171}{100}g_1^4 - \frac{9}{4}g_2^4 + \frac{63}{10} g_1^2 g_2^2 \right) 
		- \frac{3411}{2000}g_1^6 + \frac{305}{16}g_2^6 - \frac{1677}{400}g_1^4 g_2^2 - \frac{289}{80}g_1^2 g_2^4 \right],
\end{eqnarray}
 up to the two-loop level~\cite{Buttazzo:2013uya}.
We have only included the top quark Yukawa coupling,
 and omitted the other Yukawa couplings,
 since they do not contribute significantly to the Higgs quartic coupling and gauge couplings.



\end{document}